\documentclass[a4paper,,11pt]{article}

\usepackage{ifthen}
\usepackage{color}
\usepackage{epsfig}
\usepackage{amssymb}
\usepackage{graphicx}
\usepackage{amsmath}
\usepackage[center]{subfigure}

\def\be{\begin{eqnarray}}
\def\ed{\end{eqnarray}}

\newcommand{\DS}[1]{/\!\!\!#1}

\makeatletter
\def\fmslash{\@ifnextchar[{\fmsl@sh}{\fmsl@sh[0mu]}}
\def\fmsl@sh[#1]#2{%
  \mathchoice
    {\@fmsl@sh\displaystyle{#1}{#2}}%
    {\@fmsl@sh\textstyle{#1}{#2}}%
    {\@fmsl@sh\scriptstyle{#1}{#2}}%
    {\@fmsl@sh\scriptscriptstyle{#1}{#2}}}
\def\@fmsl@sh#1#2#3{\m@th\ooalign{$\hfil#1\mkern#2/\hfil$\crcr$#1#3$}}
\makeatother

\begin{document}
\begin{flushright}
\end{flushright}
\begin{center}
{\Large\bf  Subleading power corrections to  $B\to \gamma l\nu $ decay in PQCD approach }\\[2cm]
{\large\bf Yue-Long Shen$^{a}$, Zhi-Tian Zou$^{b}$ and  Yan-Bing Wei$^{d}$  }\\[0.5cm]
{\it $^a$ College of Information Science and Engineering, Ocean
University of China, Qingdao, Shandong 266100, P.R. China
 }\\
 {\it $^b$Department of Physics, Yantai University,   Yantai, Shandong 264005, P.R. China}\\
 {\it $^d$ School of Physics, Nankai University, Weijin Road 94, 300071 Tianjin, P.R. China \,}\\[1.5cm]
\end{center}
\begin{abstract}

The leptonic radiative decay $B \to \gamma l\nu$ is of great
importance in the  determination of  $B$ meson wave functions, and
evaluating the form factors $ F_{V,A}$ are the essential problem
on the study of this channel. We computed next-to-leading power
corrections to the form factors within the framework of PQCD
approach, including the power suppressed hard kernel, the
contribution from a complete set of three-particle $B$ meson wave
functions up to twist-4 and two-particle off light-cone wave
functions, the $1/m_b$  corrections in heavy quark effective
theory, and the contribution from  hadronic structure of photon.
In spite of large theoretical uncertainties, the overall
 power suppressed contributions decreases about
$50\%$ of the leading power result. The $\lambda_B$ dependence of
the integrated branching ratio is reduced after including the
subleading power contributions, thus the power corrections lead to
more ambiguity in the determination of $\lambda_B$ from $B \to
\gamma l\nu$ decay.

\end{abstract}
\vfill

\section{Introduction}

   $k_T$ factorization theorem is an appropriate theoretical
framework for exclusive $B$ meson decays.  By retaining parton
transverse momenta $k_T$, the end-point singularities which break
collinear factorization is regularized. The PQCD
approach\cite{0004004,0004213} based on the  $k_T$ factorization
framework has been applied to various exclusive processes,
especially semi-leptonic and non-leptonic $B$-meson decays, and other decay modes\cite{decay}. The
resultant predictions are in agreement with most of the
experimental data, and the most applaudable result is the CP
violation in many non-leptonic $B$-meson decay channels. The LHC-b
and forthcoming Super-B factory experiments will accumulate more
and more accurate data, which require more precise theoretical
predictions. To achieve this target, both QCD radiative
corrections and power corrections need to be considered. In PQCD
approach, QCD radiative corrections are extensively studied in
many processes, such as the pion transition form
factor\cite{Nandi:2007qx,Li:2009pr}, the pion electro-magnetic
form factors\cite{Li:2010nn,Hu:2012cp,Cheng:2014gba}, the $B \to
\pi$ form factors\cite{Li:2012nk,Cheng:2014fwa}, et al., while the
exploration  on  power corrections are very few. The motivation of
this paper is to investigate the power corrections in the leptonic
radiative decay mode $B \to \gamma l \nu$.

Most of the theoretical frameworks to study $B$ meson decays are
based on heavy quark expansion, and power corrections are
important for finite $b$ quark mass. While in the collinear
factorization, the power suppressed contributions are in general
non-factorizable due to end-point singularity, so they are often
fitted by experimental data or estimated using non-perturbative
methods.  $1/ m_b$ power corrections to $B \to \gamma \ell \nu$
were considered  at tree level \cite{Beneke:2011nf} where a
symmetry-conserving form factor $\xi(E_{\gamma})$ was introduced
to parameterize the non-local power correction.  An  approach
based on dispersion relations and quark-hadron duality was
employed to study the power suppressed contributions in $B \to
\gamma \ell\nu$ \cite{Braun:2012kp}, where the ``soft"
two-particle correction to the $B \to \gamma$ form factors was
computed at leading order. The one-loop corrections to this kind
of subleading power contribution has been computed in
\cite{Wang:2016beq}, in addition the contribution from
three-particle light-cone distribution amplitudes(LCDAs) was also
considered at tree level. In a recent paper\cite{weiyb}, using
dispersion approach, the soft contribution of power-suppressed
higher-twist corrections to the form factors that are due to
higher Fock states of $B$-meson and to the transverse momentum
(virtuality) of the light quark in the valence state was
calculated, the results are found to be much smaller than that of
twist-2 contribution. Based on the power counting in the
soft-collinear effective theory(SCET
\cite{Bauer:2000yr,Beneke:2002ph}),  the hadronic structure of
photon can contribute at next-to-leading power, which was studied
in \cite{Ball:2003fq, Wang:2018wfj}. The soft contribution and the
contribution from the hadronic structure of photon are probably
closely related, and it is interesting to uncover their
relationship.

 In the PQCD approach, tree-level power
corrections have been firstly studied in \cite{Korchemsky:1999qb},
and a more careful investigation of power corrections  was
performed in Ref.\cite{Charng:2005fj}, in which three-particle
$B$-meson wave functions, next-to-leading power(NLP) hard kernels,
and long-distance vector meson dominance contribution are
considered. In \cite{Charng:2005fj} the contribution from an
incomplete set of three-particle $B$ meson LCDAs was estimated by
power counting, but the detailed calculation is still absent. The
long distance contribution is found to be cancelled by the
radiative corrections, which makes the power correction very
small. As a rough estimate, this conclusion needs to be checked by
a more careful calculation. Our aim in this article is to make the
following improvements: (1) The contribution from a complete set
of higher twist $B$ meson wave functions, up to twist-4, will be
investigated. The higher twist wave functions include both
two-particle and three-particle Fock states, which are related by
the equation of motion.
 (2) The contribution from the hadronic structure of photon will be calculated within
 PQCD framework. As the endpoint singularity appears in the collinear factorization
is regularized by including the transverse momentum,  this kind of
contribution can be studied using factorization approach. (3) The
$1/m_b$ corrections to the heavy-to-light current in  HQET will be
considered.  Although the NLP contributions considered here are
still far from a systematical study, but they can shed light on
the correction arises from the power corrections, which makes
great sense in the determination of the parameter $\lambda_B$.

This paper is organized as follows: In the next section we will
present the analytic calculation of the decay amplitude of $B \to
\gamma l \nu $, including both  leading power(LP) and NLP
contributions. The numerical analysis is given in the third
section.   Concluding discussions are presented in Section
\ref{section: summaries}.

\section{The  $B \to \gamma l \nu$ decay amplitude  at next-to-leading power}

The radiative leptonic B-meson decay amplitude is given by
\begin{eqnarray}
 A(B\to\gamma\nu l)={G_FV_{ub}\over \sqrt{2}}\langle\gamma l\nu_l|\bar l\gamma^\nu(1-\gamma_5)\nu_l
 \bar u\gamma_\nu(1-\gamma_5) b|B\rangle.
\label{eq:Agamma}
\end{eqnarray}
At leading order in QED, the above amplitude can be written as
\begin{eqnarray}
 A(B\to\gamma\nu l)={G_FV_{ub}\over \sqrt{2}}(ig_{em}\epsilon^*_\nu)
 [T^{\nu\mu}(p,q)\bar{l}\gamma_\mu(1-\gamma_5)\nu+Q_lf_B\bar{l}\gamma_\nu(1-\gamma_5)\nu],
\label{eq:Agamma}
\end{eqnarray}
where the momenta carried by photon, lepton-pair and $B$-meson are
$p,q$ and $p+q$ respectively. In the light-cone coordinate,
$p_\mu={n\cdot p\over 2}\bar{n}_{\mu}=E_\gamma \bar{n}_{\mu}$,
$q_\mu={1\over 2}(n\cdot q \bar{n}_\mu+\bar{n}\cdot qn_\mu)$ and
$p_\mu+q_\mu=m_Bv_\mu$. The hadronic tensor $ T_{\nu\mu}$ reads
\begin{eqnarray}
 T_{\nu\mu}(p,q)=\int d^4z e^{ip\cdot z}\langle0|{\rm T}[j^{em}_\nu(z),\bar{u}(0)\gamma_\mu(1-\gamma_5)b(0)]|B(p+q)\rangle,
\end{eqnarray}
with $j^{em}_\nu(z)=\sum_q Q_q\bar{q }(z)\gamma_\nu
q(z)+Q_l\bar{l}(z)\gamma_\nu l(z)$. Considering  vector and axial
vector current conservation,  the decomposition of the hadronic
matrix element reads
\begin{eqnarray}
 T_{\nu\mu}(p,q)=-iv\cdot p\epsilon_{\mu\nu\rho\sigma}n^\rho v^\sigma
 F_V(n\cdot p)+(g_{\mu\nu}v\cdot p-v_\nu p_\mu)F_A(n\cdot
 p)+g_{\mu\nu}f_B,
\end{eqnarray}
where the last term will cancel the contribution with photon
radiated from the lepton. The differential decay rate of $B\to
\gamma l\nu$ can be readily computed using the following formula
\begin{eqnarray}
{d\Gamma\over dE_\gamma}(B \to \gamma l\nu
)={\alpha_{em}G_F^2|V_{ub}|^2\over
6\pi^2}m_BE_\gamma^3(1-{2E_\gamma\over m_B})[F_V^2(n\cdot
p)+F_A^2(n\cdot p)].
\end{eqnarray}
This equation indicates that the essential problem in the $B \to
\gamma l \nu$ decays is to study the factorization of the form
factors $F_{A,V}$. A systematical study on the power corrections
for this process needs to analyze power suppressed SCET operators,
which is rather complicated and we leave it for a future study.
Alternatively, we follow \cite{weiyb} to expand the matrix element
using heavy quark effective theory(HQET)
\begin{eqnarray}
 T_{\nu\mu}&=&\sqrt{m_B}\int d^4z e^{ip\cdot z}\langle0|{\rm T}[j^{em}_\nu(z),\bar{u}\gamma_\mu(1-\gamma_5)h_v(0)]|B(v)\rangle
 \nonumber\\
 &+&{\sqrt{m_B}\over 2m_b}\int d^4z e^{ip\cdot z}\langle0|{\rm T}[j^{em}_\nu(z),\bar{u}\gamma_\mu(1-\gamma_5)i\not\!D_\perp h_v(0)]|B(v)\rangle.
\label{hqe}\end{eqnarray}

\begin{figure}[h]
\begin{center}
\includegraphics[width=0.5  \columnwidth]{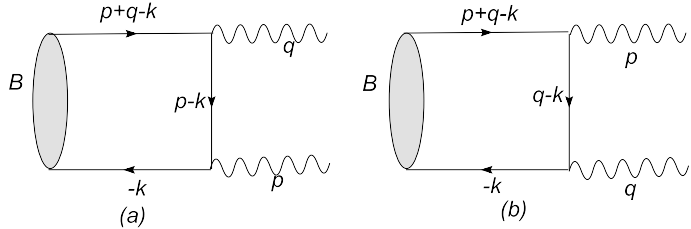}\\
 \caption{Tree level  diagrams with two-particle $B$ meson wave functions. } \label{fig: LO diagrams }
\end{center}
\end{figure}
In  the first line, the power corrections arise from the
light-cone expansion of quark propagator
\begin{equation}
\label{quarkpropagatorwith1g} \langle 0\vert
T\{q(z),\bar{q}(0)\}\vert0\rangle={i\over 2\pi^2}{\not\!z\over
z^4}-\frac{i}{16\pi^{2}} \frac{1}{z^{2}}\int^{1}_{0}d u[\not\!z
\sigma_{\alpha\beta}-4iuz_{\alpha}
\gamma_{\beta}]G^{\alpha\beta}(uz)+...,
\end{equation}
and the twist expansion of $B$-meson wave
functions\cite{CCH,CE,LRS,BS,LS,HS}
\begin{eqnarray}
\langle 0|{\bar q}(z) W_z(n)^{\dag}I_{n;y,0}W_0(n) \Gamma
h(0)|{\bar B}(v)\rangle &=& -{i f_B m_B \over 4} {\rm Tr} \bigg \{
{1 +\slash \! \! \! v \over 2} \bigg [ 2 \,
\Phi_{B}^{+}(t,z^2)+2z^2G^+(z^2)
\nonumber \\
& +& { \Phi_{B}^{-}(t,z^2)+z^2G^-(z^2)
-\Phi_{B}^{+}(t,z^2)-z^2G^+(z^2) \over t } \slash \! \! \!   z+...
\bigg ] \gamma_5 \, \Gamma\bigg \}. \label{de12p}
\end{eqnarray}
The $B$ meson wave functions describe the distributions of the
light parton in both the longitudinal direction denoted by $t=v
\cdot z$ and the transverse direction denoted by $z^2$. In the
above definition $z=(0,z^-,{\bf z}_T)$ is the coordinate of the
anti-quark field $\bar q$, $h$ is the  $b$ quark field in the
heavy quark effective theory, and $\Gamma$ represents a Dirac
matrix. The Wilson line
 $W_z(n)$ is written as
\begin{eqnarray}
\label{eq:WL.def} W_z(n) = { P} \exp\left[-ig \int_0^\infty
d\lambda n\cdot A(z+\lambda n)\right].
\end{eqnarray}
 The vertical link $I_{n;z,0}$ at
infinity does not contribute in the covariant gauge \cite{CS08}.
Due to the light-cone divergences associated with the Wilson
lines, the light-cone vector should be rotated to satisfy
$n^2\not=0$. The wave functions $\Phi_{B}^{\pm}(t,z^2)$ are
twist-2, and $G^\pm(z^2)$ are twist-4. In addition, the definition
of three-particle LCDAs is as follows
\begin{eqnarray}
\langle 0|\bar{q}_{2\alpha}(z)G_{\mu\nu}(uz) b_{\beta}(0)
|\bar{B}_v\rangle
 &= &\frac{f_B m_B}{4}\int _0^\infty d\omega\int_0^\infty d\xi \int{d^2k_{1\perp}\over(2\pi)^2}
  \int{d^2k_{2\perp}\over (2\pi)^2}
 e^{-i(k_{1}+uk_{2})\cdot z}\nonumber
\\&\times &\bigg [(1 +\DS v)
\bigg \{(v_\mu\gamma_\nu-v_\nu\gamma_\mu) [\psi_A- \psi_V] -
i\sigma_{\mu\nu}\psi_V-(\bar n_\mu v_\nu-\bar n_\nu v_\mu)X_A
\nonumber
\\ &+&(\bar n_\mu \gamma_\nu-\bar n_\nu \gamma_\mu)(W+Y_A)+
i\epsilon_{\mu\nu\rho\sigma}\bar n^\rho v^\sigma
\gamma_5\tilde{X}-i\epsilon_{\mu\nu\rho\sigma}\bar n^\rho
\gamma^\sigma \gamma_5\tilde{Y} \nonumber
\\ &-&
(\bar n_\mu v_\nu-\bar n_\nu v_\mu)\not\! \bar n W+(\bar n_\mu
\gamma_\nu-\bar n_\nu \gamma_\mu)\not\! \bar n Z \bigg
\}\gamma_5\bigg]_{\beta\alpha}
(\omega,\xi,k_{1\perp},k_{2\perp})\label{del3p}.
\end{eqnarray}
The wave functions defined above do not have definite twist, but
they are convenient in the calculation for their simple Lorentz
structure.

For the second line of Eq.(\ref{hqe}), although there already
exists a suppressed factor $1/m_b$, higher twist $B$ meson wave
functions are still required as the power expansion in terms of
$1/m_b$ is not equivalent to the twist expansion. In the following
we will consider the contribution from leading twist and higher
twist $B$ meson wave functions respectively in the first line of
Eq.(\ref{hqe}), and then evaluate the contribution from the second
line Eq.(\ref{hqe}). Furthermore, we will also investigate the
contribution from the hadronic structure of photon at the last
subsection.
\subsection{Contribution from leading twist $B$ meson wave functions}
Firstly we consier LP result  of $F_{V,A}$ and NLP corrections
from leading twist $B$ meson wave functions. From the definition
in Eq.(\ref{de12p}), the momentum space projector for  $B$-meson
twist-2 wave functions can be written by
\begin{eqnarray}
\label{bproj} M^B_{\alpha\beta} =- \frac{i f_B m_B}{4}\Bigg[
\frac{1+\not \! v}{2} \Bigg\{ \phi^+_B(\omega,k_\perp)\,\not \! n
+ \phi^-_B(\omega,k_\perp)\,\not \! \bar n
 - \,\int_0^{\omega} d\eta \,\left(\phi^-_B(\eta,k_\perp)-
\phi^+_B(\eta,k_\perp)\right)\,\gamma^\mu\frac{\partial}{\partial
k_{\perp\mu}} \Bigg\} \, \gamma_5
\Bigg]_{\alpha\beta}.\label{EQ.bwave}
\end{eqnarray}
The leading power contribution is from Fig.(\ref{fig: LO diagrams
}a), in which the light quark propagator can be decomposed as
\begin{eqnarray}
{i(\not \! p-\not \! k)\over (p-k)^2}=-i{E_\gamma \not\! \bar
n\over 2E_\gamma \omega+k_\perp^2}+i{\omega \not\!  v\over
2E_\gamma \omega+k_\perp^2}-i{\not \! k_\perp\over 2E_\gamma
\omega+k_\perp^2},\label{eq:propagator}
\end{eqnarray}
where the first term  is at leading power, and the other two terms
are suppressed by $\lambda={\omega\over E_\gamma}$. Taking only
the leading power contribution into account, the form factors
$F_{V,A}$ can be written by
\begin{eqnarray}
F_A^{LP}(E_\gamma)=F_V^{LP}(E_\gamma)={2\over 3} f_Bm_B\int_0^1
d\omega \int {d^2k_\perp \over
(2\pi)^2}{\phi^+_B(\omega,k_\perp)\over 2\omega E_\gamma
+k_\perp^2}.\label{eq:lp1}
\end{eqnarray}
According to \cite{Charng:2005fj}, the mass dependence of the
hadron state arises if the power suppressed operators $O_{1,2}$
are included
\begin{eqnarray}
\langle 0|{\bar u}_{\rho}(z)h_{\delta}(0)|\bar
B(Mv)\rangle_{QCD}=\sum_{i=1,2}\langle 0|i\int d^4yT[{\bar
u}_\rho(z) h_{\delta}(0)O_{i}(y)]|\bar B(v)\rangle\;.\label{m4}
\end{eqnarray}
where
\begin{eqnarray}
O_1=\frac{1}{m_b}\bar h(iD)^2h\;,\;\;\;O_2=\frac{g}{2m_b}\bar
h\sigma^{\mu\nu}G_{\mu\nu}h\;.\label{o12}
\end{eqnarray}
After considering the mass dependence of the hadronic state the
momentum fraction of the soft quark inside the $B$ meson can be
defined by $x=\omega/ m_B$, and Eq.(\ref{eq:lp1}) turns to
\begin{eqnarray}
F_A^{LP}(E_\gamma)=F_V^{LP}(E_\gamma)={2\over 3} f_Bm_B\int_0^1 dx
\int {d^2k_\perp \over (2\pi)^2}{\phi^+_B(x,k_\perp)\over 2xm_ B
E_\gamma +k_\perp^2}.
\end{eqnarray}
 The QCD correction to  the  $B$ meson wave
functions and the leading-order (LO) hard kernel produces both the
single and double logarithms  $\ln^2 {k_\perp\over E_\gamma}$,
$\ln {k_\perp\over E_\gamma}$,  $\ln^2 x$ and $\ln x$
respectively, which become large as $k_\perp \ll E_\gamma, x\ll
1$. These large logarithms need to be resummed, among them $k_T$
resummation leads to Sudakov form factor, and threshold
resummation(resumming $\ln^2 x$ and $\ln x$) leads to jet
function. The $k_T$ and threshold resummation improves the
convergence of the perturbation series, and the resummation
improved factorization formula can be rewritten by
\begin{eqnarray}
F_A^{LP}(E_\gamma)=F_V^{LP}(E_\gamma)={2\over 3} f_Bm_B\int_0^1 dx
\int_0^\infty bdb K_0(\sqrt{2xE_\gamma
m_B}b)S_t(x)e^{-s_B(t)}\phi^+_B(x,b),
\end{eqnarray}
where $s_B(t)$ is the Sudakov form factor and $S_t(x)$ is the jet
function from the threshold resummation\cite{0004004,0004213}. The
threshold factor from the resummation of $\ln^2 x$ has been
parameterized as
\begin{eqnarray}
S_t(x,Q) &=&{2^{1+c(Q^2)} \, \Gamma ({3 \over 2}+c(Q^2))  \over
\sqrt{\pi} \, \Gamma(1+c(Q^2)) } \, \left [ x (1-x) \right
]^{c(Q^2)}   \,.
\end{eqnarray}
Both the hard kernel and the wave function have been transformed
into the impact parameter space($b$ space) because it is more
convenient to perform sudakov resummation in $b$ space. In the
above equation the resummation of rapidity logarithms $\ln
{(n\cdot p)^2\over n^2}$, which will cause scheme dependence, is
neglected. In \cite{Li:2012md} the joint resummation with respect
to all the large logarithms is performed, and this effect will be
considered in the future study.

The power suppressed amplitude includes the latter two terms in
Eq.(\ref{eq:propagator}) and the contribution from Fig(\ref{fig:
LO diagrams }b). We note that the last term in
Eq.(\ref{eq:propagator}), which is related to the transverse
derivative in the $B$-meson wave function(the last term in
Eq.(\ref{EQ.bwave})), vanishes in 4-dimension due to the Lorentz
structure
$\gamma_{\perp\mu}\not\!\epsilon^*_\perp\gamma_\perp^\mu$. The
second term in Eq.(\ref{eq:propagator}) results in
\begin{eqnarray}
F_A^{NP1a}(E_\gamma)&=&-{1\over 3E_\gamma} f_Bm_B^2\int_0^1 dxx
\int_0^\infty bdb S_t(x)e^{-s_B(t)}K_0(\sqrt{2xE_\gamma
m_B}b)[\phi^-_B(x,b)+\phi^+_B(x,b)]\nonumber \\
F_V^{NP1a}(E_\gamma)&=&{1\over 3E_\gamma} f_Bm_B^2\int_0^1 dxx
\int_0^\infty bdb S_t(x)e^{-s_B(t)}K_0(\sqrt{2xE_\gamma
m_B}b)[\phi^-_B(x,b)-\phi^+_B(x,b)].
\end{eqnarray}
 The internal line in  Fig(\ref{fig: LO
diagrams }b) is a heavy quark propagator, due to the basic idea of
effective theory, it must be integrated out and leads to local
contribution. In the diagrammatic approach, the propagator is
proportional to $1\over 2m_bE_\gamma+k_\perp^2$, where $k_\perp^2$
in the denominator is obviously suppressed, and this term is
identical to the collinear factorization result after $k_\perp$ is
dropped£º
\begin{eqnarray}
F_A^{NLP1b}(E_\gamma)=-F_V^{NLP1b}(E_\gamma)={ f_Bm_B\over
6m_bE_\gamma}
\end{eqnarray}
Adding up the leading twist NLP contribution, we obtain
\begin{eqnarray}
F_{A,V}^{NLP1}(E_\gamma)=F_{A,V}^{NLP1a}(E_\gamma)+F_{A,V}^{NLP1b}(E_\gamma).
\end{eqnarray}

\subsection{Contribution from higher twist $B$ meson wave functions}

Up to twist-4, the higher twist $B$ meson wave functions include
two-particle Fock state, i.e. $G^\pm(t,z^2)$, and three-particle
Fock state defined in Eq.(\ref{del3p}). According to twist
expansion, the three-particle wave functions include one twist-3,
$\phi_3=\psi_A-\psi_V$, and three twist-4,
$\phi_4=\psi_A+\psi_V,\,\psi_4=\psi_A+X_A,\,\tilde\psi_4=\psi_V-\tilde
X_A$, in which only two wave functions are independent.  We assume
that all the wave functions have the factorized form, i.e.,
$G^\pm(t,z^2)=G^\pm_B(t)\tilde \Sigma(z^2)$, where $G^\pm_B(t)$
are $B$-meson LCDAs. The two-particle and three-particle LCDAs are
related by the following equation of motion
\begin{equation}
2t^{2}G^{+}_B(t)=-\frac{1}{2}\Phi^{-}_{B}(t)-\bigg(t\frac{d}{d
t}-\frac{1}{2}+it\bar
{\Lambda}\bigg)\Phi^{+}_{B}(t)-t^{2}\int^{1}_{0}d
u\bar{u}\Psi_{4}(t,ut),
\end{equation}
and it is convenient to define
\begin{equation}
2t^{2}\hat{G}^{+}_B(t)=-\frac{1}{2}\Phi^{-}_{B}(t)-\bigg(t\frac{d}{d
t}-\frac{1}{2}+it\bar {\Lambda}\bigg)\Phi^{+}_{B}(t).
\end{equation}

\begin{figure}[t]
\begin{center}
\includegraphics[width=0.25  \columnwidth]{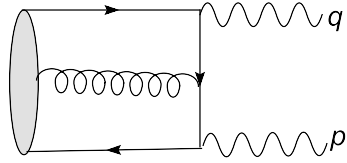}\\
 \caption{Diagrams of the contribution from three-particle $B$ meson wave functions} \label{fig: 3particle}
\end{center}
\end{figure}

The contribution from three-particle Fock state is plotted in
Fig.(\ref{fig: 3particle}). Inserting Eq.(\ref{del3p}) and
Eq.(\ref{quarkpropagatorwith1g}) into the correlation function
$T_{\nu\mu}$, one can obtain the factorization formulae of
contributions from three-particle $B$-meson wave functions.
Combining the three-particle contribution with the contribution
from $G^\pm(t,z^2)$, we have

\begin{eqnarray}
F_{A}^{NLP2a}(E_\gamma)=F_{V}^{NLP2a}(E_\gamma) &= & -{4\over 3}
f_Bm_B\int_0^\infty {d\omega\over\sqrt{2E_\gamma \omega }}
\int^\infty_0b^2db \hat{g}_B^+(\omega,b)K_1(\sqrt{2E_\gamma \omega
}b),
\\
F_{A}^{NLP2b}(E_\gamma)=F_{V}^{NLP2b}(E_\gamma) &=& {1\over
3\sqrt{2E_\gamma}} f_Bm_B\int_0^\infty d\omega\int_0^\infty
d\xi\int_0^1 {du\over \sqrt{\omega+u\xi} }\int_0^\infty
b^2db\nonumber
\\ &\times&K_1(\sqrt{2E_\gamma (\omega+u\xi)}b) \,
\left[\psi_4(\omega,\xi,b) -\tilde{\psi}_4(\omega,\xi,b)\right],
\end{eqnarray}
where $\hat G^+({t,b})=\int_0^\infty d\omega e^{-i\omega t}\hat
g^+({\omega},b)$.  The total contribution from high twist wave
functions is written by
\begin{eqnarray}
F_{A,V}^{NLP2}(E_\gamma)=F_{A,V}^{NLP2a}(E_\gamma)+F_{A,V}^{NLP2b}(E_\gamma).
\end{eqnarray}

\subsection{Power suppressed contribution in HQET}
To evaluate the $1/m_b$ correction in Eq.(\ref{hqe}), one should
take advantage of the formula
\begin{eqnarray}
 \bar q(z)\Gamma D_\rho h_v(0)=\partial_\rho[\bar q(z)\Gamma
 h_v(0)]+i\int_0^1du\bar u\bar q(z)z^\lambda G_{\lambda\rho}(zu)\Gamma
 h_v(0)+[{\partial\over \partial z^\rho }\bar q(z)]\Gamma
 h_v(0).
\end{eqnarray}
For the first term in the above equation, using the following
relation
\begin{eqnarray}
 \langle0|\partial_\rho[\bar q(z)\Gamma
 h_v(0)]|B(v)\rangle={\partial\over \partial y^\rho}e^{-i\bar \Lambda v\cdot y}\langle0|[\bar q(z)\Gamma
 h_v(0)]|B(v)\rangle=-i\bar\Lambda v_\rho\langle0|[\bar q(z)\Gamma
 h_v(0)]|B(v)\rangle,
\end{eqnarray}
one can obtain
\begin{eqnarray}
F_A^{NLP3a}(E_\gamma)=F_V^{NLP3a}(E_\gamma)={\bar \Lambda\over
3m_b} f_Bm_B\int_0^1 dx \int_0^\infty bdb K_0(\sqrt{2xE_\gamma
m_B}b)S_t(x)e^{-s_B(t)}\phi^+_B(x,b).
\end{eqnarray}
The matrix element of  the second term is related to the twist-3
three-particle wave function, following the same method with the
above subsection, we have
\begin{eqnarray}
F_A^{NLP3b}(E_\gamma)=F_V^{NLP3b}(E_\gamma) &=&
{\sqrt{2E_\gamma}\over 3m_b} f_Bm_B\int_0^\infty
d\omega\int_0^\infty d\xi\int_0^1 {du\over \sqrt{\omega+u\xi}
}\int_0^\infty b^2db\nonumber
\\ &\times&
  K_1(\sqrt{2E_\gamma
(\omega+u\xi)}b)\phi_3(\omega,\xi,b).\end{eqnarray}
 The last term
can be evaluated with integration by part, and the result reads
\begin{eqnarray}
F_A^{NLP3c}(E_\gamma)=F_V^{NLP3c}(E_\gamma) &=&{f_Bm_B\over 3m_b}
\int_0^1 dx \int_0^\infty bdb K_0(\sqrt{2xE_\gamma
m_B}b)\Psi_B(x,b)\nonumber
\\ &-&{2f_Bm_BE_\gamma\over 3m_b}
\int_0^1 xdx \int_0^\infty bdb K_0(\sqrt{2xE_\gamma
m_B}b)\phi^+_B(x,b),
\end{eqnarray}
with $\Psi_B(\omega,b)=\int_0^\omega
d\eta[\phi^-(\eta,b)-\phi^+(\eta,b)]$. Adding up all the above
results we have
\begin{eqnarray}
F_{A,V}^{NLP3}(E_\gamma)=F_{A,V}^{NLP3a}(E_\gamma)+F_{A,V}^{NLP3b}(E_\gamma)+F_{A,V}^{NLP3c}(E_\gamma).\label{heqt}
\end{eqnarray}
\subsection{Contribution from hadronic structure of photon}
\begin{figure}[t]
\begin{center}
\includegraphics[width=0.5  \columnwidth]{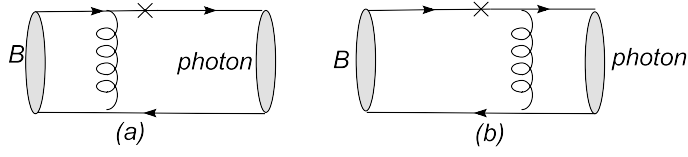}\\
 \caption{Diagrams of the contribution from hadronic structure of photon } \label{fig: photon}
\end{center}
\end{figure}

To investigate the contribution  of the hadronic structure of
photon, it is essential to introduce the LCDAs of photon, which
have been studied up to twist-4 level in \cite{Ball:2002ps}. In the
present paper we will only consider the contribution of two-particle
twist-2 and twist-3 LCDAs, which are defined below
\begin{eqnarray}
\langle \gamma(p,\lambda)|\bar
 q(z)\sigma_{\alpha\beta}q(0)|0\rangle
&=&ig_{em}Q_q\langle\bar
qq\rangle(p_\beta\epsilon^*_\alpha-p_\alpha\epsilon^*_\beta)\int_0^1
   du e^{iup\cdot z}[\chi(\mu)\phi_\gamma(u,\mu)]
\nonumber \\
\langle \gamma(p,\lambda)|\bar
 q(z)\gamma_{\alpha}q(0)|0\rangle
&=&-g_{em}Q_qf_{3\gamma}\epsilon^*_\alpha\int_0^1
   du e^{iup\cdot z}\psi^{(v)}_\gamma(u,\mu)
\nonumber \\
\langle \gamma(p,\lambda)|\bar
 q(z)\gamma_{\alpha}\gamma_5q(0)|0\rangle
&=&{1\over
4}g_{em}Q_qf_{3\gamma}\epsilon_{\alpha\beta\rho\sigma}p^\rho
z^\sigma\epsilon^{*\beta}\int_0^1
   du e^{iup\cdot z}\psi^{(a)}_\gamma(u,\mu)
\end{eqnarray}\\
where $\phi_\gamma(u,\mu)$ is twist-2, and
$\psi^{(a,v)}_\gamma(u,\mu)$ are twist-3. The  normalization
constants of  these LCDAs depend on the factorization scale, and
the evolution behavior is written by
\begin{eqnarray}
 \chi(\mu) &=& \left[{\alpha(\mu)\over
 \alpha_s(\mu_0)}\right]^{16\over 33-2n_f}\chi(\mu_0),\;\;\;\langle\bar q q\rangle(\mu)=\left[{\alpha(\mu_0)\over
 \alpha_s(\mu)}\right]^{12\over 33-2n_f}\langle\bar q
 q\rangle(\mu_0),\label{sd1}
\end{eqnarray}
\begin{eqnarray}
&& f_{3 \gamma}(\mu)= \left [  {\alpha_s (\mu) \over \alpha_s
(\mu_0)} \right ]^{23  \over 99-6n_f} \,
 f_{3 \gamma}(\mu_0)\,.  \qquad\label{sd2}
\end{eqnarray}
In the factorization formulae we will neglect the transverse
momentum dependence of the wave functions, because the Sudakov
effect for light $q\bar q$ state is significant, and further
suppression is not necessary. The momentum space projector for the
two-particle LCDAs is written by(up to two-particle twist-3)
\begin{eqnarray}
M^\gamma_{\alpha\beta} &=&{1\over 4}g_{em}Q_q\bigg\{-
\langle\bar
qq\rangle(\not\!\epsilon^*\not\! p)\chi(\mu)\phi_\gamma(u,\mu)
-f_{3\gamma}(\not\!\epsilon^*)\psi^{(v)}_\gamma(u,\mu)
\nonumber \\
 &-&{i\over 8}f_{3\gamma}\epsilon_{\mu\nu\rho\sigma}(\gamma^{\mu}\gamma^5){\bar n}^\rho
\epsilon^{*\nu}[n^\sigma{d\over
du}\psi^{(a)}_\gamma(u,\mu)-2E_\gamma\psi^{(a)}_\gamma(u,\mu){\partial\over
\partial k_{\perp\sigma}}]\bigg\}_{\alpha\beta}.
\end{eqnarray}
The matrix element of $B \to \gamma$ transition can be calculated
through the convolution formula
\begin{eqnarray}
_{HS}\langle \gamma|\bar q\Gamma b|B\rangle={4\pi\alpha_s C_F\over
N_c}\int_0^1 dx\int_0^\infty b_1db_1\int_0^1 du\int_0^\infty
b_2db_2M^B_{\beta\rho}H^\Gamma_{\alpha\beta\rho\sigma}M^\gamma_{\sigma\alpha},
\end{eqnarray}
after evaluating the Feynman diagrams in Fig(\ref{fig: photon}),
the results of the form factors $F_{V,A}$ read
\begin{eqnarray}
F_A^{NLP4a}(E_\gamma) &=&-{\pi\alpha_sC_Ff_Bm_BQ_u\over E_\gamma
N_c}\int_0^1 dx\int_0^\infty b_1db_1\int_0^1 du\int_0^\infty
b_2db_2
h^a_e(x,u,b_1,b_2)\nonumber \\
&\times&[2 E_\gamma m_B\langle\bar
qq\rangle\chi(\mu)\phi_\gamma(u,\mu)\phi_B^--(2uE_\gamma\phi_B^++m_B(\phi_B^-+\phi_B^+))
f_{3\gamma}\psi^{(v)}\nonumber \\
&+&{1\over 2}(uE_\gamma\phi_B^++{1\over
2}m_B(\phi_B^--\phi_B^+))f_{3\gamma}{d\over
du}\psi^{(a)}_\gamma(u,\mu)],
\end{eqnarray}
\begin{eqnarray}
F_V^{NLP4a}(E_\gamma) &=&-{\pi\alpha_sC_Ff_Bm_BQ_u\over E_\gamma
N_c}\int_0^1 dx\int_0^\infty b_1db_1\int_0^1 du\int_0^\infty
b_2db_2
h^a_e(x,u,b_1,b_2)\nonumber \\
&\times&[2E_\gamma m_B\langle\bar
qq\rangle\chi(\mu)\phi_\gamma(u,\mu)\phi_B^-+(2uE_\gamma\phi_B^+
+m_B(\phi_B^--\phi_B^+))f_{3\gamma}\psi^{(v)}\nonumber \\
&-&{1\over 2}(uE_\gamma\phi_B^++{1\over
2}m_B(\phi_B^-+\phi_B^+))f_{3\gamma}{d\over
du}\psi^{(a)}_\gamma(u,\mu)],
\end{eqnarray}

\begin{eqnarray}
F_A^{NLP4b}(E_\gamma) &=&{\pi\alpha_sC_Ff_Bm_BQ_u\over E_\gamma
N_c}\int_0^1 dx\int_0^\infty b_1db_1\int_0^1 du\int_0^\infty
b_2db_2
h^b_e(x,u,b_1,b_2)\nonumber \\
&\times&[2E_\gamma\phi_B^+f_{3\gamma}\psi^{(v)}+{1\over
2}E_\gamma\phi_B^+f_{3\gamma}{d\over du}\psi^{(a)}_\gamma(u,\mu)],\\
F_V^{NLP4b}(E_\gamma) &=&F_A^{NLP4b}(E_\gamma),
\end{eqnarray}
with the hard functions
\begin{eqnarray}
h^a_e(x,u,b_1,b_2)&=&e^{-s_B(t)-s_\gamma(t)}\left[\theta(b_1-b_2)I_0(\sqrt
{2uE_\gamma m_B}b_2)K_0(\sqrt {2uE_\gamma
m_B}b_1)\right.\nonumber\\
&& \left.+\theta(b_2-b_1)I_0(\sqrt {2uE_\gamma m_B}b_1)K_0(\sqrt
{2uE_\gamma
m_B}b_2)\right]K_0(\sqrt {2xum_B E_\gamma}b_1)S_t(u),\nonumber\\
h^b_e(x,u,b_1,b_2)&=&e^{-s_B(t)-s_\gamma(t)}\left[\theta(b_1-b_2)I_0(\sqrt
{2xE_\gamma m_B}b_2)K_0(\sqrt {2xE_\gamma
m_B}b_1)\right.\nonumber \\
&& \left.+\theta(b_2-b_1)I_0(\sqrt {2xE_\gamma m_B}b_1)K_0(\sqrt
{2xE_\gamma m_B}b_2)\right]K_0(\sqrt {2xum_B E_\gamma}b_1)S_t(u).
\end{eqnarray}
Summing up the two diagrams, the form factors from photon hadronic
structure can be written by
\begin{eqnarray}
F_{V}^{NLP4}(E_\gamma)&=&F_{V}^{NLP4a}(E_\gamma)+F_{V}^{NLP4b}(E_\gamma),\nonumber
\\
F_{A}^{NLP4}(E_\gamma)&=&F_{A}^{NLP4a}(E_\gamma)+F_{A}^{NLP4b}(E_\gamma).
\end{eqnarray}

In summary, combining all the NLP contributions together, we have
\begin{eqnarray}
F_{V}^{NLP}(E_\gamma)&=&F_{V}^{NLP1}(E_\gamma)+F_{V}^{NLP2}(E_\gamma)+F_{V}^{NLP3}(E_\gamma)+F_{V}^{NLP4}(E_\gamma),\nonumber
\\
F_{A}^{NLP}(E_\gamma)&=&F_{A}^{NLP1}(E_\gamma)+F_{A}^{NLP2}(E_\gamma)+F_{A}^{NLP3}(E_\gamma)+F_{A}^{NLP4}(E_\gamma).
\end{eqnarray}
Based on the calculations in above sections, several comments are
as follows:
\begin{itemize}
\item{All the results of the form factors are given at tree level.
The radiative corrections are of great importance in the hard
exclusive processes, and in the $B\to \gamma l\nu$ decay it can
reduce the leading order amplitude by $20\%-25\%$ in collinear
factorization\cite{DescotesGenon:2002mw,Lunghi:2002ju,Bosch:2003fc,Beneke:2011nf}.
In $k_T$ factorization, the NLO corrections have been studied in
\cite{Korchemsky:1999qb}, while the endpoint behavior in this
study is under controversy, and a more comprehensive study is
required, which is left for a future study.}
 \item{For the
contributions from higher twist $B$ meson wave functions up to
twist-4, it has been found that it is free from endpoint
singularity in the collinear factorization, thus the endpoint
region is not very important and the Sudakov form factor and jet
function is not essential. In addition, there is no study on the
$k_T$ resummation effect for the higher twist wave functions so
far, so the Sudakov factor is not considered here. If
four-particle twist-5 and twist-6 wave functions are included,
there does exist endpoint singularity\cite{weiyb}, and the
resummation effect must be considered.} \item{Only two-particle
twist-2 and twist-3 photon LCDAs are employed in the contribution
from the hadronic structure of photon. In \cite{Wang:2018wfj}, the
contributions from the full set of photon LCDAs up to twist-4 are
studied using light-cone sum rules approach, and the results
indicate that the contribution from two-particle twist-2 LCDA is
dominant, and the contribution from higher twist and
three-particle LCDAs is suppressed. Here we neglect higher twist
photon LCDAs except for two-particle twist-3 contribution which
gives important contributions to the $B\to V$ form factors in the
PQCD approach.}
\end{itemize}

\section{Numerical analysis}
The most important input  parameters are  wave functions of $B$
meson and  photon. We have assumed that the transverse momentum
dependent $B$ meson wave function ${\phi}_{B}^{\pm} (x,k_T)$
possesses the factorized form
\begin{eqnarray}
\phi_{B}^{\pm}(x,k_T) =  \phi_{B}^{\pm}(x) \, \Sigma(k_T) \,,
\end{eqnarray}
where the transverse part needs to be transformed into the impact
parameter space through Fourier transform, and the wave functions
turns to
\begin{eqnarray}
\tilde\phi_{B}^{\pm}(x,b) =  \phi_{B}^{\pm}(x) \,
\tilde{\Sigma}(b) \,.
\end{eqnarray} For the transverse part the Gaussian model
is usually adopt in the PQCD approach, i.e.
$\tilde{\Sigma}(b)=e^{-{1\over 2}\omega_0^2b^2}$, with
$\omega_0=\lambda_B$. For the longitudinal part,  we employ the
following models  to check the model dependence of the form
factors. The first one is a free parton
model\cite{Kawamura:2001jm}
\begin{eqnarray}
\phi_{B\rm I}^{+}(x) &=& { x \over 2 x_1^2} \, \theta(2
x_1-x),\nonumber \\
\phi_{B\rm I}^{-} (x)&=& { 2 x_1 -x \over 2 x_1^2} \, \theta(2
x_1-x),
\end{eqnarray}
where $x_1=x_0=\omega_0/m_B$. The second one is from the QCD sum
rules with local duality approximation\cite{Braun:2003wx}
\begin{eqnarray}
\phi_{B\rm II}^{+}(x)&=& { 3x \over 4 x_2^3} \, (2 x_2-x)\theta(2 x_2-x),\nonumber\\
\phi_{B\rm II}^{-}(x) &=& { 1 \over
8x_2^3}[3(2x_2-x)^2+{10(\lambda_E^2-\lambda_H^2)\over
3x_2^2m_B^2}(3x^2-6xx_2+2x_2^2)] \, \theta(2 x_2-x),
\end{eqnarray}
where $ x_2=3/2x_0$. In the phenomenological studies with PQCD
approach, a more widely used model is as follows,
\begin{eqnarray}
\phi_{B\rm III}^{+}(x)&=& \phi_{B\rm
III}^{-}(x)=N_Bx^2(1-x)^2e^{-{x^2\over 2x_0^2}},
\end{eqnarray}
where normalization constant $N_B$ is determined by $\lambda_B$.
For the model of two-particle twist-4 $B$-meson LCDA, following
\cite{weiyb} we adopt
\begin{eqnarray}
   \hat{g}^+_B(\omega)&=&{\omega^2\over 2\lambda_B}
   \bigg(1-{\lambda_E^2-\lambda_H^2\over 36\lambda_B^2}\bigg)e^{-{\omega\over \lambda_B}},
\end{eqnarray}
where the parameters $\lambda_E^2$ and $\lambda_H^2$ which are related to
the matrix element of local quark-gluon operator can be estimated
with QCD sum rules approach.
  The three-particle $B$ wave function is also supposed to satisfy $\psi(\omega,\xi,b,
  ub)=\psi(\omega,\xi)\tilde\Sigma(b)$, and
 the exponential model of the longitudinal part is widely used
\begin{eqnarray}
    \phi_3(\omega,\xi)&=&{\lambda_E^2-\lambda_H^2\over 6\lambda_B^5}\omega\xi^2 e^{-{\omega+\xi\over \lambda_B}},\nonumber \\
    \psi_4(\omega,\xi)&=&{\lambda_E^2\over 3\lambda_B^4}\omega\xi e^{-{\omega+\xi\over \lambda_B}},\nonumber \\
   \tilde\psi_4(\omega,\xi)&=&{\lambda_H^2\over 3\lambda_B^4}\omega\xi e^{-{\omega+\xi\over \lambda_B}}.
\end{eqnarray}

The  light-cone distribution amplitudes $\phi_{\gamma}(u),
\psi^{(v,a)}(\omega,\xi)$ have been systematically studied in
Ref.\cite{Ball:2002ps}, and the expressions are quoted as follows.
The two particle twist-2 LCDA is expanded in terms of Gegenbauer
polynomials,
\begin{eqnarray}
    \phi_\gamma(u,\mu)  &=& 6u\bar u
  \left[1+\sum_{n=2}^\infty b_n(\mu_0)
  C^{3/2}_n(u-\bar u)\right],
\end{eqnarray}
and twist-3 LCDAs in conformal expansion read
\begin{eqnarray}
\psi^{(v)}(\xi, \mu) &=& 5 \, \left ( 3\, \xi^2 -1  \right ) + {3
\over 64} \, \left [ 15 \, \omega_{\gamma}^{V}(\mu) - 5\,
\omega_{\gamma}^{A}(\mu)  \right ] \,
\left ( 3 - 30 \, \xi^2 +  35 \,  \xi^4 \right ) \,,  \nonumber \\
\psi^{(a)}(\xi, \mu) &=&  {5 \over 2} \, \left ( 1 -  \xi^2 \right
) \, (5 \, \xi^2 -1 ) \left ( 1 + {9 \over 16} \,
\omega_{\gamma}^{V}(\mu) - {3 \over 16} \,
\omega_{\gamma}^{A}(\mu)  \right ) \,.
 \end{eqnarray}
 In addition to the normalization
constant(\ref{sd1},\ref{sd2}), the scale dependence of the
parameters in the LCDAs can be written as
\begin{eqnarray}
&& b_2(\mu)= \left [ {\alpha_s (\mu) \over \alpha_s (\mu_0)}
\right ]^{8  \over  33-2n_f} \,
 b_2(\mu_0)\,,  \qquad
\nonumber \\
&& \left(
\begin{array}{c}
\omega^{V}_\gamma(\mu) - \omega^{A}_\gamma(\mu) \\
\omega^{V}_\gamma(\mu) + \omega^{A}_\gamma(\mu)
\end{array}
\right) = \left (  {\alpha_s (\mu) \over \alpha_s (\mu_0)} \right
)^{\Gamma_{\omega}  / \beta_0} \, \left(
\begin{array}{c}
\omega^{V}_\gamma(\mu_0) - \omega^{A}_\gamma(\mu_0) \\
\omega^{V}_\gamma(\mu_0) + \omega^{A}_\gamma(\mu_0)
\end{array}
\right) \,,
\end{eqnarray}
where the anomalous dimension matrix $\Gamma_{\omega}$  and
$\beta_0$ is given by \cite{Ball:2002ps,Ball:1998sk}
\begin{eqnarray}
\Gamma_{\omega}&=& \left(
\begin{array}{c}
3 \, C_F - {2 \over 3} \, C_A  \qquad   {2 \over 3} \, C_F - {2 \over 3} \, C_A   \\
{5 \over 3} \, C_F - {4 \over 3} \, C_A \qquad  {1 \over 2} \, C_F
+ C_A
\end{array}
\right)\,,\nonumber \\
\beta_0&=&11-{2\over 3}n_f.
\end{eqnarray}
The  value of the parameters used in the calculations are
presented in Table(\ref{para}), among them the scale dependent
parameters are given at $\mu_0=1.0GeV$. These parameters should be
run to the factorization scale $t$ in numerical analysis.

\begin{table}
\centerline{\parbox{14cm}{\caption{\label{para} Numerical value of
the parameters entering the calculations}}} \vspace{0.1cm}
\begin{center}
\begin{tabular}{c|c|c|c|c}
\hline\hline parameter &$\omega_0(\lambda_B)$ &
$x_0$&$\chi(1GeV)$ & $\langle\bar qq\rangle(1GeV)$ \\
\hline value &$0.35\pm0.10GeV$ & $0.076\pm0.015$&$(3.15\pm
0.03)GeV^{-2}$ &$-[(256^{+14}_{-16})MeV]^3$
\\
\hline\hline parameter &$b_2(1GeV)$ &
$f_{3\gamma}(1GeV)$&$\omega^V_\gamma(1GeV)$ & $\omega^A_\gamma(1GeV)$ \\
\hline value &$0.07\pm0.07$ & $ -(4 \pm 2)
 \times 10^{-3}\,{\rm GeV}^2$ &$3.8\pm 1.8$&$-2.1 \pm 1.0$
\\
\hline\hline  parameter&$N_B$ &$\lambda_E^2$ &$\lambda_H^2$
&$f_B$\\
\hline value &$3417$ &$0.06\pm0.04GeV^2$&$0.12\pm0.05GeV^2$&$0.19\pm0.02GeV$\\
\hline\hline
\end{tabular}
\end{center}
\vspace{-0.2cm}
\end{table}

Now we present the numerical results for the form factors
$F_{V,A}$ and the branching ratio of $B\to \gamma\nu l$ decay. In
physical interesting photon energy region
$1.5GeV<E_\gamma<2.6GeV$, the leading power results of
$F_{V,A}(E_\gamma)$ at tree level are plotted in Fig(\ref{fig:
lp}), where all the parameters are fixed at the central values in
Table({\ref{para}}). At leading power $F_V=F_A$ due to the
left-handness of the standard model. The three curves are from the
three models of leading twist $B$ meson wave functions, and the
difference between them is only about $3\%-5\%$. In the following
 we set the model $\phi^\pm_{B\rm III}$ as default, which approaches zero
 at endpoint region.  $\phi^-_{B\rm I,II}$ does not vanish when $x=0$,
 and it will lead to too large endpoint contribution when entering the factorization formula.
 Compared with the result of leading order $F_{V,A}$ in collinear factorization,
 the PQCD result is relatively
smaller  due to  the inclusion of transverse momentum in the
denominator of the propagators as  well as suppression from $k_T$
resummation and threshold resummation.

\begin{figure}[t]
\begin{center}
\includegraphics[width=0.45  \columnwidth]{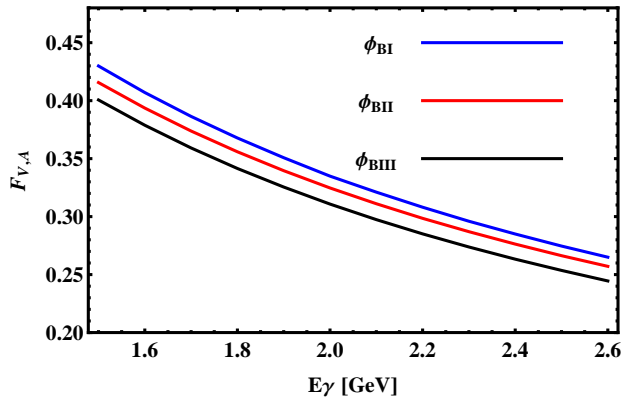}\\
 \caption{The leading power contribution to the form factors $F_{V,A}$, where blue,red and black curves
  are corresponding to the wave functions $\phi_{B\rm I}$,$\phi_{B\rm II}$ and $\phi_{B\rm III}$ respectively. }\label{fig: lp}
\end{center}
\end{figure}

\begin{figure}[t]
\begin{center}
\includegraphics[width=0.45
\columnwidth]{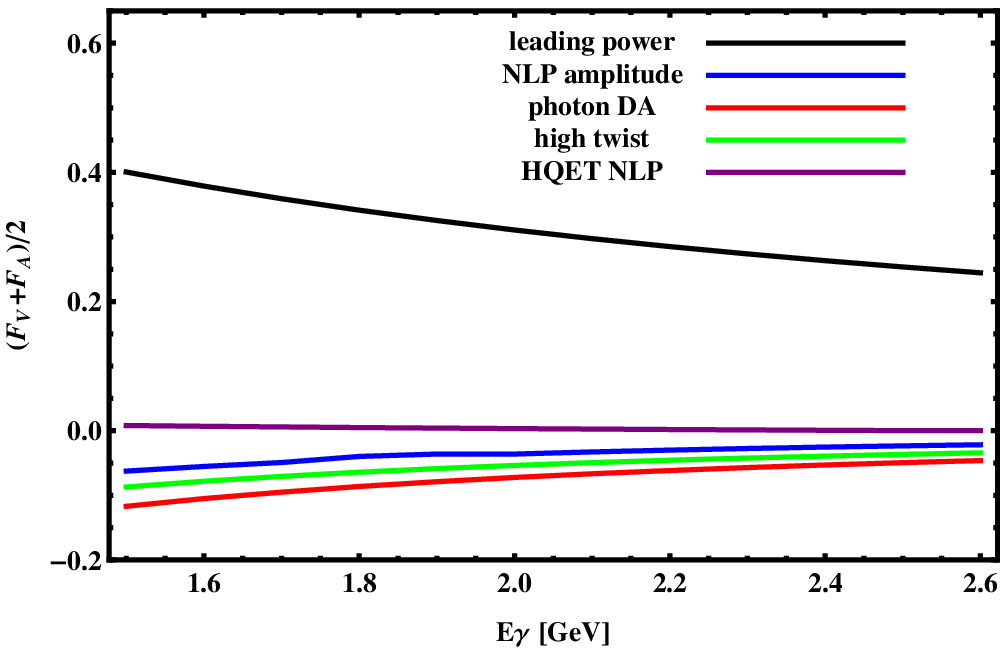}
\includegraphics[width=0.45  \columnwidth]{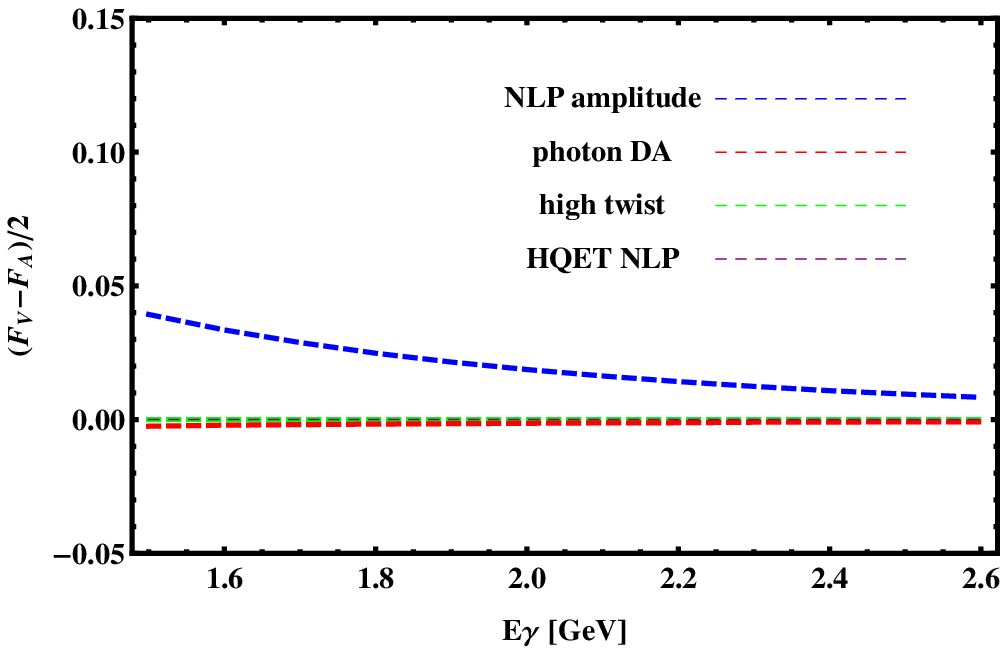}\\
 \caption{The next-to-leading power contribution to the form factors $F_{V,A}$.
  The left(right) panel denotes the photon momentum
 dependence of $(F_V+F_A)/2$($(F_V-F_A)/2$)respectively.  } \label{fig:nlp}
\end{center}
\end{figure}

The NLP contribution to the form factors are presented in
Fig(\ref{fig:nlp}). Among various kinds of contributions, which
from hadronic structure of photon is most important. It decreases
the leading power contribution by about $20\%$ for the symmetric
form factor $(F_V+F_A)/2 $, and this result is consistent with the
predictions from light-cone sum rules\cite{Wang:2018wfj}. It can
only give rise to a minor contribution to the symmetry breaking
part $(F_V-F_A)/2 $  because the leading twist photon LCDA
provides identical result for $F_V$ and $F_A$, and the symmetry
breaking effect  is only from higher twist photon LCDA . The
contribution from higher twist $B$-meson wave functions, including
both two-particle and three-particle Fock states, also decreases
the leading power contribution by about $20\%$, and it keeps the
symmetry between $F_V$ and $F_A$. The contribution from three
particle $B$-meson  wave functions is much smaller than that from
higher twist two-particle wave function, which is consistent with
the rough estimate in\cite{Charng:2005fj}. The power suppressed
hard kernel
 can also give rise to sizeable corrections as
the suppression factors $\omega/E_\gamma$ is not very small when
$E_\gamma$ is not large. It is the main source of symmetry
breaking part $(F_V-F_A)/2 $.  The $1/m_b$ suppression term from
HQET is negligible due to the cancellation between different part
in Eq.(\ref{heqt}). The different pieces  of the NLP corrections
considered in this paper are all sizable except for the $1/m_b$
suppression term from HQET, furthermore, the effects of them are
all negative. The overall NLP correction is then significant, it
decreases the LP result by about $50\%$. This result indicates the
extraordinarily importance of  power corrections in this channel.

\begin{figure}[t]
\begin{center}
\includegraphics[width=0.45  \columnwidth]{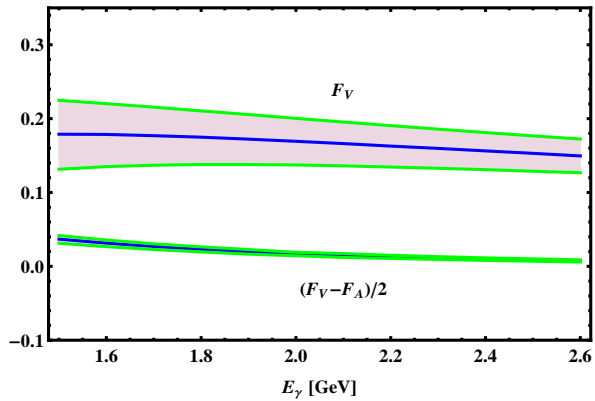}\\
 \caption{The form factors with uncertainty } \label{fig: uncertainty}
\end{center}
\end{figure}

Now we present the uncertainties from the various parameters in
Table(\ref{para}). If we fix $E_\gamma=2.0GeV$ and
$\lambda_B=0.35GeV$, then the form factors with uncertainty are
obtained as(in the unit of GeV)
\begin{eqnarray}
F_V(2GeV)=0.169&+&\bigg(^{+0.003}_{-0.003}\bigg)_{\lambda_E^2-\lambda_H^2}
+\bigg(^{+0.020}_{-0.020}\bigg)_{f_{3\gamma}}+\bigg(^{+0.003}_{-0.003}\bigg)_{b_2}
+\bigg(^{+0.011}_{-0.011}\bigg)_{\omega^V_\gamma}
+\bigg(^{+0.002}_{-0.002}\bigg)_{\omega^A_\gamma}\nonumber\\
&+&\bigg(^{+0.018}_{-0.018}\bigg)_{f_B}+\bigg(^{+0.006}_{-0.006}\bigg)_{\langle\bar
qq\rangle}+\bigg(^{+0.009}_{-0.012}\bigg)_{S_t}
\\F_A(2GeV)=0.135&+&\bigg(^{+0.003}_{-0.003}\bigg)_{\lambda_E^2-\lambda_H^2}
+\bigg(^{+0.018}_{-0.019}\bigg)_{f_{3\gamma}}+\bigg(^{+0.003}_{-0.003}\bigg)_{b_2}
+\bigg(^{+0.008}_{-0.008}\bigg)_{\omega^V_\gamma}
+\bigg(^{+0.001}_{-0.002}\bigg)_{\omega^A_\gamma}\nonumber\\
&+&\bigg(^{+0.015}_{-0.014}\bigg)_{f_B}+\bigg(^{+0.006}_{-0.006}\bigg)_{\langle\bar
qq\rangle}+\bigg(^{+0.008}_{-0.011}\bigg)_{S_t}
\end{eqnarray}
where the important sources of the uncertainties include the
parameters $f_{3\gamma}$ and $\omega_ V$ in the distribution
amplitude of photon, the decay constant of $B$ meson, and the
parameter $c$ in the threshold resummation. For simplicity,
$c(Q^2)$ has been fixed as a constant and varies in the region
$[0.45,0.65]$. Due to the variation regions of the twist-2
parameters $\chi(\mu_0)$ and $\langle \bar qq\rangle$ are very
small,  the uncertainties from them are not important. The
$E_\gamma$ dependence of the form factors with uncertainties is
plotted in Fig.(\ref{fig: uncertainty}), where the errors are
added in quadrature, and the overall uncertainty is expressed in
the shaded region. Here the form factor $F_A$ is not shown for its
uncertainty region is overlapped with $F_V$, instead, the
uncertainty region of the symmetry breaking effect $(F_V-F_A)/2$
is presented. The uncertainty region of $F_V$ is large because the
parameters in the $B$ meson and photon wave functions are not well
determined, and they should be constrained by more preciously
measured physical quantities such as $B \to \pi$ transition form
factors.

\begin{figure}
\begin{center}
\includegraphics[width=0.45 \columnwidth]{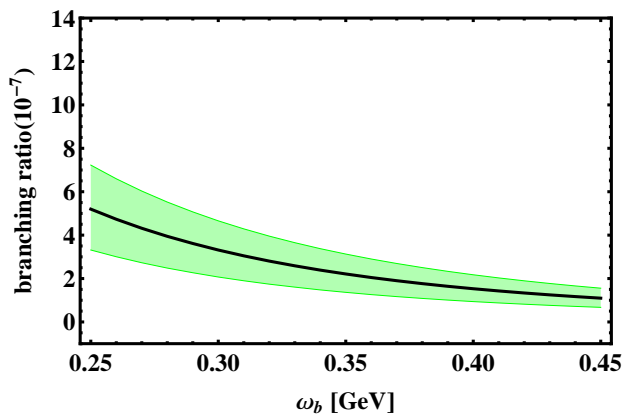}
\caption{Dependence of the partial branching fractions ${\cal
BR}(B \to \gamma \ell \nu, \, E_{\gamma} \geq E_{\rm cut})$ on the
first inverse moment $\lambda_B(\mu_0)$ for $E_{\rm cut}=1.5 \,
{\rm GeV}$ (blue band) and $E_{\rm cut}=2.0 \, {\rm GeV}$ (green
band).} \label{fig: predicted branching ratio}
\end{center}
\end{figure}

 Having the theoretical predictions of the form factors $F_{V,A}$ in our hands,
 we proceed to discuss the theory
constraints on the first inverse moment $\lambda_B$ using
integrated branching ratios of $B \to \gamma\nu l$.  The lower
limit of integral should be a photon-energy cut to get rid of the
soft photon radiation. The integrated branching fractions with the
phase-space cut on the photon energy read
\begin{eqnarray}
{\cal BR}(B \to \gamma \ell \nu, \, E_{\gamma} \geq E_{\rm cut}) =
\tau_B \, \int_{E_{\rm cut}}^{m_B/2} \, d E_{\gamma} \, {d \,
\Gamma(B \to \gamma \ell \nu) \over d \, E_{\gamma}} \,,
\end{eqnarray}
where $\tau_B$ indicates the lifetime of the $B$-meson. Our
predictions for the partial branching ratios of  $B \to \gamma
\ell \nu$ decay including power suppressed contributions are
displayed in Fig.(\ref{fig: predicted branching ratio}). The
variation range of the first inverse moment $\lambda_B$  is
$[0.25, 0.45] \, {\rm GeV}$. It can be observed that the
integrated branching fractions ${\cal BR}(B \to \gamma \ell \nu,
\, E_{\gamma} \geq E_{\rm cut})$ grow   with the decrease of
$\lambda_B$, but the slope becomes small then $\lambda_B$ is
getting large, in addition, the theoretical uncertainty is big.
This $\lambda_B$ dependence behavior makes it more difficult to
preciously determine the parameter $\lambda_B$. Recently, Belle
collaboration reported their improved measurement of  the
branching ratio of $B \to \gamma \ell \nu$ with the energy cut
$E_\gamma> 1$GeV\cite{Gelb:2018end}, the measured branching ratio
is given by
\begin{eqnarray} {\cal
BR}(B \to \gamma \ell \nu, \, E_{\gamma} \geq {\rm 1.0GeV}) =
(1.4\pm1.0\pm0.4)\times 10^{-6} \,,
\end{eqnarray}
and a Bayesian upper  limit of ${\cal BR}(B \to \gamma \ell \nu,
\, E_{\gamma} \geq {\rm 1.0GeV}) < 3.0\times 10^{-6} \,$ is
determined at 90\% confidence level. Furthermore, the predictions
and uncertainties of partial decay rate in  Ref.\cite{weiyb}
extrapolated to $E_\gamma > 1$ GeV are used to determine
$\lambda_B$. While if our result is employed, the uncertainty of
$\lambda_B$ determined from $B \to \gamma \ell \nu$ decay should
be larger. Thus a more systematic study of the NLP corrections to
this channel is of great importance. On the experimental side,  it
is meaningful to measure the branching fraction with the
phase-space cut on the photon energy larger than 1.5GeV, which is
helpful to reduce model dependence.

\section{CONCLUSION AND DISCUSSION}

\label{section: summaries}

The leptonic radiative decay $B \to \gamma l\nu$ is believed to be
an ideal channel to determine the $B$ meson wave functions,
especially the first inverse moment $1/\lambda_B$, which is an
important input in the semi-leptonic and non-leptonic $B$ meson
decays.  In the study of $B \to \gamma l\nu$ decay, the key
problem is to investigate the form factors $F_{V,A}(E_\gamma)$. We
computed next-to-leading power corrections to the form factors
within the framework of PQCD approach, including the power
suppressed hard kernel, the contribution from a complete set of
three-particle $B$ meson wave functions up to twist-4 and two
particle off light-cone wave functions, the $1/m_b$  corrections
in HQET, and the contribution from the hadronic structure of
photon taking advantage of two-particle twist-2 and twist-3 photon
LCDAs. In the study of  power corrections, PQCD approach has its
unique advantage because it is free from endpoint singularity
through keeping transverse momentum of parton. Numerically, both
the contribution from the higher twist $B$ meson wave functions
and the hadronic structure of photon can reduce the leading power
result by about $20\%$, and the power suppressed hard kernel
decrease the leading power amplitude over $10\%$. The overall
results is about $50\%$ smaller than leading power, under the
condition that the QCD radiative corrections are not considered.
Within the parameter space in this paper, the power correction is
so important that one can hardly using the leading power result to
reasonably determine the $B$ meson wave function.  After including
the power corrections, the integrated branching ratio of $B \to
\gamma \nu l$ grows with decreasing $\lambda_B$, but the rate of
change is smaller than the leading power case, in addition to the
large theoretical uncertainty, it is difficult to preciously
determine $\lambda_B$ only employing this processes.  We should
point out that our study is far from a systematic investigation,
and more efforts need to be made to uncover the influence of the
power corrections. With more and more precise measurements of
$B\to \gamma l \nu$ decay, the parameters in $B$ meson wave
functions must be better constrained.

\section*{Acknowledgement}
We are grateful to H. N. Li for useful discussions and comments.
This work was supported in part by National Natural Science
Foundation of China under the Grants Nos. 11705159, 11447032; and the
Natural Science Foundation of Shandong province under the
Grant No. ZR2018JL001.
 \vspace{0.5 cm}


\end{document}